\documentclass[epj]{svjour}

\usepackage{graphicx}
\usepackage{bm}
\usepackage{amsmath}
\usepackage{amssymb}
\usepackage[active]{srcltx}

\newcommand{\beq}{\begin{equation}}
\newcommand{\eeq}{\end{equation}}
\newcommand{\bea}{\begin{eqnarray}}
\newcommand{\eea}{\end{eqnarray}}

\begin{document}
\title{\hfill{\small HISKP-TH-08/07}\\[0.5cm]
%
A Model Study of Discrete Scale Invariance and Long-Range Interactions}

\author{H.-W. Hammer\inst{1} and R. Higa\inst{1}}

\institute{
Helmholtz-Institut f\"ur Strahlen- und Kernphysik (Theorie),
Universit\"at Bonn, Nu\ss allee 14-16, 53115 Bonn, Germany}

\date{Received: date / Revised version: date}

\abstract{
We investigate the modification of discrete scale invariance 
in the bound state spectrum by 
long-range interactions. This problem is relevant for effective field theory 
descriptions of nuclear cluster states and manifestations of the
Efimov effect in nuclei. 
As a model system, we choose a one dimensional
inverse square potential supplemented with a long-range Coulomb
interaction. We study the renormalization and bound-state spectrum 
of the system as a function of the Coulomb interaction strength.
Our results indicate, that the counterterm required to renormalize the
inverse square potential alone is sufficient to renormalize the full
problem. However, the breaking of the discrete scale invariance 
through the Coulomb interaction leads to a modified bound state spectrum.
The shallow bound states are strongly influenced by the Coulomb interaction 
while the deep bound states are dominated by the inverse square 
potential.
}

\PACS{{11.10.Gh}      {Renormalization} \and 
{03.65.-w}      {Quantum Mechanics} \and 
{05.10.Cc}      {Renormalization Group methods} 
}

\authorrunning{H.-W. Hammer and R. Higa}
\titlerunning{Discrete Scale Invariance and Long-Range Interactions}
\maketitle

\section{Introduction}

The application of effective field theory (EFT) methods to nuclear physics
is by now well established \cite{Bedaque:2002mn}.
If there is a separation of scales
in a physical system, effective field theory allows for
controlled calculations of low-energy observables with well-defined
error estimates. 
In nuclear physics, mostly effective field theories with 
nucleons and pions (and possibly Deltas) as degrees of freedom are used. 
However, for a certain class of systems, it is possible to use effective
field theories with even more effective degrees of freedom
\cite{Braaten:2004rn}. This is the realm of halo nuclei and nuclear 
cluster states.

A halo nucleus is one that consists of a tightly bound core 
surrounded by one or more loosely bound valence nucleons.
The valence nucleons are characterized by a very low
separation energy compared to those in the core. 
As a consequence, the radius of the halo nucleus is large 
compared to the radius of the core.
The separation of scales in halo nuclei leads to universal properties
that are insensitive to the structure of the core (see, e.g., 
Ref.~\cite{DFHT00} and references therein). The most
carefully studied Borromean halo nuclei are $^6$He and $^{11}$Li, 
which have two weakly bound valence neutrons \cite{ZDFJV93,JRFG04}. 
In the case of $^6$He, the core is an alpha particle.
An EFT framework to describe halo systems was introduced
in \cite{BHvK1,BHvK2}
where the neutron-alpha ($n\alpha$) system was studied
in an EFT with nucleon and alpha degrees of freedom. 
Further extensions to the proton-alpha and alpha-alpha systems
were considered in Refs.~\cite{BRvK,Higa:2008dn}.
Similar concepts can be applied to nuclear cluster states.
The best-known example is the 
structure of $^{12}$C. This system has an excited $0^+$ state, the 
so-called Hoyle state which shows a clear clustering into three
$\alpha$ particles. This observation suggests that this state 
can be described by an EFT of $\alpha$ particles inteacting via
short-range contact interactions.
An important question is whether there is an universal binding mechanism
for these systems. A prime 
candidate for such a mechanism is the Efimov effect 
\cite{Efimov70}. 

In an EFT framework, the Efimov effect can be 
related to a renormalization group (RG) limit cycle \cite{Albe-81}.
Most applications of the RG involve a flow towards a fixed point, where the 
system is scale invariant. However, as 
pointed out by Wilson \cite{Wilson:1970ag}, one can also have closed 
curves under the RG flow in the space of coupling constants. The RG
flow completes a cycle around the curve every time the cutoff is
changed by a multiplicative factor $\lambda_0$. This number $\lambda_0$
is the preferred scaling factor. A necessary
condition for a limit cycle is invariance under discrete scale
transformations: $x\to \lambda_0^n x$, where $n$ is an integer. This
discrete scaling symmetry is reflected in log-periodic behavior of
physical observables.
The Efimov effect can be understood as the manifestation of a 
limit cycle in the bound state spectrum of the three-body problem
with large scattering length $a$. This limit cycle property is 
manifest in the EFT treatment of Refs.~\cite{Bedaque:1998kg},
where an explicit log-periodic three-body counterterm is introduced.
In the limit $a \to \pm \infty$, there is an accumulation of
3-body bound states near threshold with binding energies differing by
multiplicative factors of $\lambda_0^2 \simeq 515.03$ \cite{Efi71}.
Recently, the first convincing experimental evidence for this effect 
was obtained by measuring its effect on three-body loss rates in
a gas of cold Cs atoms \cite{Grimm06}.

The Efimov effect could also be responsible for the binding 
of certain halo nuclei and cluster states
\cite{Federov:1994cf}.  In particular for the 
latter, however, the short-range strong interaction is usually 
accompanied by a long-range Cou\-lomb interaction.
The effect of such 
long-range Coulomb interactions on the physics of limit cycles 
and discrete scale invariance is therefore an important issue. 
In order to get some insight into this question,
we start with a simpler problem
that has also a limit cycle behavior: the
one-dimensional Schr\"odinger equation with an attractive 
inverse square potential. If the attraction is larger than
a certain critical value, the system also shows a limit cycle.
This limit cycle becomes evident in a bound state spectrum with discrete
scale invariance similar to the Efimov effect.
Indeed, the inverse square potential is intimitely connected to
the three-body system with large scattering length, which
reduces to a one-dimensional Schr\"odinger equation with an 
inverse square potential in the hyperradius for large momenta
\cite{Efi71,Braaten:2004rn}.
This makes the inverse square potential an ideal model system to 
study the physics of limit cycles and discrete scale invariance
\cite{Beane:2000wh,Bawin:2003dm,Braaten:2004pg,Barford:2004fz,Hammer:2005sa,long-2007}.

In this paper, we study the effect of 
a long-range Cou\-lomb interaction on discrete scale invariance 
in the bound state spectrum
for an inverse square potential with a Cou\-lomb potential.
(For an earlier study of the interplay between Coulomb and strong
interactions in exotic atoms, see Ref.~\cite{Gal96}.)
In the next section, we briefly review the renormalization of the inverse 
square potential in the approach of Ref.~\cite{Hammer:2005sa}.
In Sec.~\ref{sec:longrange}, we introduce the long-range Coulomb potential.
The renormalization and our results for the bound state spectrum
are discussed in Sec.~\ref{sec:renorm} and in Sec.~\ref{sec:pert}
a perturbative treatment of the Coulomb interaction is given.
Finally, we present our conclusions in Sec.~\ref{sec:conc}.
Our treatment of the Coulomb divergence is described in 
Appendix \ref{sec:CBdiv}.

\section{Inverse Square Potential}
\label{sec:1or2}

In order to set up our problem, we briefly review the renormalization
of the $1/r^2$ potential in momentum space without the long-range
Coulomb interaction \cite{Hammer:2005sa}.
We consider the attractive inverse square potential
\beq
V_S(r) = \frac{\hbar^2}{m}\frac{c}{r^2}\,,
\qquad\mbox{with}\quad r\equiv|\vec{r}|\,,\quad
c\equiv -\frac{1}{4}-\nu^2\,,
\label{eq:defpot}
\eeq
and $\nu > 0$ a positive real parameter.
This potential has the same scaling behavior as the kinetic energy 
operator and, consequently, is scale invariant at the classical level.
In the following, we set the particle
mass and Planck's constant $m=\hbar=1$ for convenience.
For values of $c\geq -\frac{1}{4}$, the potential is well-behaved and
the corresponding Schr\"odinger equation has a unique solution, 
see Ref.~\cite{FLS71}. However,
we are interested in the case $c< -\frac{1}{4}$ which corresponds
to real values of $\nu$ in (\ref{eq:defpot}).
In this case, the $1/r^2$ potential is singular and the
usual boundary conditions for the Schr\"{o}dinger equation do not
lead to a unique solution. 
We can calculate the Fourier transform of the potential using
dimensional regularization (see Ref.~\cite{Hammer:2005sa} for details). 
This leads to the expression 
\beq 
V_S(q)=\frac{2\pi^2 c}{q}\,
\label{eq:FT1oR2} 
\eeq 
for the momentum space representation of the $1/r^2$ potential 
(\ref{eq:defpot}). Since the potential is local, its Fourier transform 
depends only on the momentum transfer $q$.

\begin{figure}[ht]
\centerline{\includegraphics*[width=8cm,angle=0]{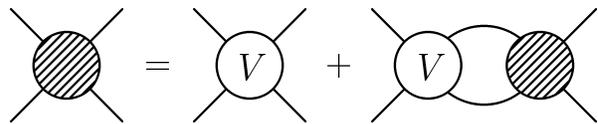}}
\caption{Lippmann-Schwinger equation for a two-body potential $V$.}
\label{fig:LS}
\end{figure}

The Lippmann-Schwinger (LS) equation for two particles interacting
via $V_S(q)$ from Eq.~(\ref{eq:FT1oR2}) in their center-of-mass frame
takes the form
\beq
T_E (\vec{p},\vec{p}') = V_S(|\vec{p}-\vec{p}'|)
+\int \frac{d^3 q}{(2\pi)^3}
   \frac{V_S(|\vec{p}-\vec{q}|)}{E-q^2+i\epsilon}
    \,T_E(\vec{q} ,\vec{p}')\,,
\label{eq:LSeqraw}
\eeq
where $E$ is the total energy and
$\vec{p}$ ($\vec{p}'$) are the relative momenta of the incoming (outgoing)
particles, respectively. A pictorial representation of this equation is
given in Fig.~\ref{fig:LS}.
We are only interested in the S-wave contribution. In higher partial 
waves the singular behavior of the potential is screened by the angular 
momentum barrier, but for sufficiently strong attraction it will become
visible as well (see, e.g., Ref.~\cite{long-2007}). 
Projecting onto S-waves by
integrating the equation over the relative angle between $\vec{p}$ and
$\vec{p}'$: $\frac{1}{2}\, \int d\cos\theta_{\vec{p}\vec{p}'}$,
we obtain the integral equation
\bea
t_E (p,p')&=&v_S(p,p')
+ \frac{1}{2\pi^2}\int_0^\infty \frac{dq\,q^2}{E-q^2+i\epsilon}
\nonumber\\ && 
\times\,v_S(p,q)\,t_E (q,p')\,,
\label{eq:LSeq}
\eea
where
\beq
v_S(p,q)=2\pi^2 c\left(\frac{\theta(p-q)}{p}+\frac{\theta(q-p)}{q}\right)\,.
\eeq

The physical observables are the bound state spectrum
and the scattering phase shifts $\delta(k)$. 
The phase shifts are determined by the
solution to Eq.~(\ref{eq:LSeq}) evaluated at the on-shell point $E=k^2$,
$k=p'=p$ via
$t_{k^2}(k,k)=-4\pi/(k\cot\delta(k)-ik)$\,.
Since $p'$ appears only as a parameter in Eq.~(\ref{eq:LSeq}),
we can set $p'=p$ to simplify the equation.
The binding energies are given by those values of $E<0$ for which
the homogeneous version of Eq.~(\ref{eq:LSeq}) has a solution.
For the bound state equation the dependence of the solution $\phi_E(p)$
on $p'$ disappears altogether.

It is well-known that Eq.~(\ref{eq:LSeq}) does not have a unique solution 
since the $1/r^2$ potential for real $\nu$ is singular \cite{FLS71}.
The most general solution of the bound state equation
for $E=0$ can be written as
\beq
\phi_0 (p)={\cal N}\, p^{-1/2} \left(p^{i\nu}e^{i\varphi}+
   p^{-i\nu}e^{-i\varphi}\right)\,,
\label{eq:2sol}
\eeq
where the relative phase $\varphi$ is a free parameter. 
The value of $\varphi$ is not determined by the $1/r^2$ 
potential and has to be taken from elsewhere. It is exactly this phase
$\varphi$ which is fixed by self-adjoint extensions of the
potential \cite{Bawin:2003dm}.

In the framework of an effective theory, 
this is conveniently done using renormalization theory.
We regularize the LS equation by applying a momentum cutoff
$\Lambda$ and include a momentum-independent counterterm
$\delta V_S(\Lambda)$ in the potential.
The precise form of the cutoff, for example Gaussian cutoff or sharp
cutoff, is not important, but we use a sharp cutoff for simplicity.
Making the replacement
\beq
V_S(q) \quad \Rightarrow \quad
V_S(q)+\delta V_S (\Lambda)= \frac{2\pi^2 c}{q} + 
\frac{H_S(\Lambda)}{\Lambda}\,
\eeq
in Eq.~(\ref{eq:LSeqraw})
the LS equation (\ref{eq:LSeq}) 
for bound state solutions with $E=-E_B<0$ becomes
\begin{eqnarray}
\phi_{E_B} (p)&=& 
-\frac{1}{2\pi^2}\int_0^\Lambda \frac{dq\,q^2}{E_B+q^2}
\nonumber\\[1mm]&&\times
\left[v_S(p,q)+\frac{H_S(\Lambda)}{\Lambda}\right]\,\phi_{E_B} (q)\,.
\label{eq:LSeqCT}
\end{eqnarray}
The functional dependence of $H_S(\Lambda)$ can be determined
analytically from invariance of low-energy observables 
under renormalization group transformations. 
We demand that the relative phase
of the zero-energy bound solution of Eq.~(\ref{eq:LSeqCT})
remains unchanged under variations of the cutoff $\Lambda$ and find
\beq
H_S(\Lambda)=2\pi^2 c \,\frac{1-2\nu\tan(\nu\ln(\Lambda/\Lambda_*))}
         {1+2\nu\tan(\nu\ln(\Lambda/\Lambda_*))}\,,
\label{eq:Hdep}
\eeq
where $\Lambda_*$ is a free parameter that determines the relative
phase in (\ref{eq:2sol}): $\varphi=-\nu\ln\Lambda_*$. 
In order to fix $\varphi$, 
we can either specify both the cutoff $\Lambda$ and the dimensionless 
coupling $H$ or, using Eq.~(\ref{eq:Hdep}), one dimensionful parameter: 
$\Lambda_*$. This parameter $\Lambda_*$ is generated by the iteration of
quantum corrections in solving the integral equation  (\ref{eq:LSeqCT}).
This is similar to the phenomenon of dimensional transmutation
in QCD \cite{Wil99}.

Note that  $H_S(\Lambda)$ remains unchanged when the argument is multiplied 
by $\lambda_0^{n}$, where $n$ is an integer number and $\lambda_0
=e^{\pi/\nu}$ is the discrete scaling factor. This discrete scaling
symmetry is a consequence of the limit cycle and 
reflects itself in physical observables 
\cite{Braaten:2004rn,Hammer:2005sa}. In the
bound state spectrum, for example, the ratio of consecutive binding energies 
is a constant, $E_{n}/E_{n+1}=\lambda_0^{2\pi/\nu}$. 
Therefore, we can study modifications of the limit cycle through 
modifications to the discrete scaling symmetry of the bound state
spectrum.

Moreover, the discrete symmetry implies the existence of a set of cutoffs
\beq
\Lambda_n(\Lambda_*)=\Lambda_* \exp(n\pi/\nu)\,
\label{eq:lambdaN}
\eeq
with $H_S(\Lambda_n)\equiv 0$. 
We can therefore obtain a renormalized version of Eq.~(\ref{eq:LSeqCT})
that does not explicitly contain the counterterm by using
the discrete set of cutoffs from Eq.~(\ref{eq:lambdaN}).
The same trick can be used for the three-body problem with large
scattering length \cite{Hammer:2000nf}.

\section{Inclusion of the Long-Range Potential}
\label{sec:longrange}

We now include an additional attractive Coulomb potential of the 
form
\beq
V_C(q)=-\frac{4\pi \alpha}{q^2+\gamma^2}\,,
\eeq
where $\gamma$ is a photon mass that will be taken to zero in the
end. The speed of light has been set to unity for convenience.
In the following, we vary the strength $\alpha$ of the potential. 
Projecting onto S-waves as discussed in the previous
section we find
\beq
v_C(p,q)=-\frac{\pi\alpha}{pq}\ln\left[
         \frac{(p+q)^2+\gamma^2}{(p-q)^2+\gamma^2}\right]\,.
\label{eq:vc}
\eeq
The integral equation for bound state solutions, Eq.~(\ref{eq:LSeqCT}),
then becomes
\bea
\phi_{E_B} (p)&=& 
-\frac{1}{2\pi^2}\int_0^\Lambda \frac{dq\,q^2}{E_B+q^2}
\,\bigg[v_S(p,q)
\nonumber\\ && \quad
+\frac{H_S(\Lambda)}{\Lambda}
+v_C(p,q)\bigg]\,\phi_{E_B} (q)\,.
\label{eq:LSeqCTCb}
\eea

From Eq.~(\ref{eq:vc}), it is clear that the kernel of 
the integral equation (\ref{eq:LSeqCTCb}) 
diverges for $q=p$ in the limit $\gamma\to 0$:
this is the well-known Coulomb singularity. Its 
origin can be traced back to the integral equation for the scattering 
amplitude with a long-range Coulomb interaction, which diverges at 
forward angles. When projected into S-waves, this singularity appears
in the diagonal terms of the potential in momentum space. 
For the binding energies this singularity should not be 
a problem, as long as the diagonal terms are handled properly. 
In  Appendix~\ref{app}, we describe how to treat these terms, 
based on the the idea outlined in Ref.~\cite{Coultrick}.

\section{Renormalization and Results}
\label{sec:renorm}

First we address the renormalization of the full problem
including Coulomb. It is 
not clear a priori whether the introduction of the Coulomb potential
will require an additional counterterm. 

In order to answer this 
\begin{figure*}[ht]
\centerline{\includegraphics*[height=6cm]{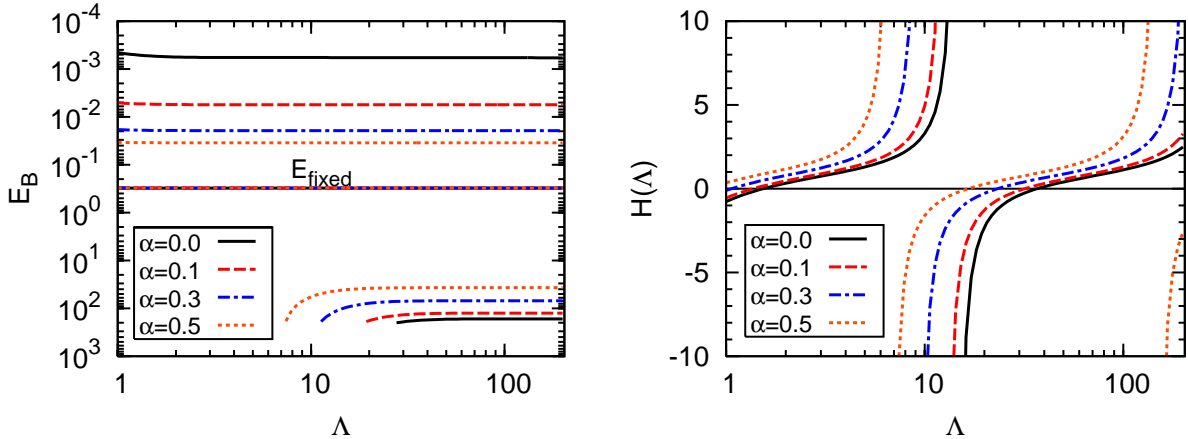}}
\caption{Bound state spectrum with one state fixed at $E_B\approx 0.308$ 
(left panel) and the counterterm required for renormalization 
(right panel) for $\alpha=0, 0.1, 0.3, 0.5$ and with $\nu=\Lambda_*=1$, 
as a function of $\Lambda$. 
In the right panel, the curve $\alpha=0$ coincides with 
$H(\Lambda)$ given by Eq.~(\ref{eq:Hdep}), except for $\Lambda\lesssim 1$.
}
\label{fig:plot2}
\end{figure*}
question, we calculate the bound state spectrum at a given cutoff
and choose to fix the binding energy of one bound state
as the cutoff $\Lambda$ is varied.
The fixed energy was chosen as one of the pure $1/r^2$ states, 
$E_{\rm fixed}\approx 0.308$.\footnote{With our choice of units all
energies and momenta are dimensionless.} 
We then calculate the bound state spectrum for other values of the cutoff
$\Lambda$ and adjust the counterterm $H_S(\Lambda)$ numerically to
keep this binding energy fixed. The results of this analysis are
shown in Fig.~\ref{fig:plot2}. The left panel shows the 
bound states for $\alpha=0, 0.1, 0.3, 0.5$
as a function of $\Lambda$. 
In order to keep this and the remaining figures as legible as possible, 
only the three deepest states are shown. 
The case $\alpha=0$ corresponds to the 
pure $1/r^2$ potential. One observes
that the cutoff dependence of all states is removed once the 
counterterm is adjusted to fix one of the states. 
Note, however, that the deepest state shows a cutoff dependence near the 
cutoff where it first appears with infinite binding energy. 
This behavior has nothing to do with the long-range interaction and is due 
to the way the system is renormalized, keeping low-energy physics unchanged.
A similar behavior is also observed for the pure $1/r^2$ potential
and the three-body system with large scattering length
\cite{Bedaque:1998kg,Braaten:2004pg,Hammer:2005sa}.
 
Moreover, it is evident that the long-range interaction destroys
the discrete scale invariance in the spectrum. Only for $\alpha=0$,
the ratio of consecutive binding energies is a constant.
In the right 
panel of Fig.~\ref{fig:plot2}, we show the numerically determined 
values of the counterterm $H_S(\Lambda)$ as a function of $\Lambda$
for  $\alpha=0, 0.1, 0.3, 0.5$. The figure suggests that the long
range potential shifts the argument of the counterterm 
$H_S(\ln\Lambda) \to H_S(\ln\Lambda
+f(\alpha))$ where $f(\alpha)$ is a monotonic function of the coupling
$\alpha$. This implies that the long-range Coulomb 
potential merely renormalizes the value of $\Lambda_*$. As a consequence,
the counterterm $H_S(\Lambda)$ from Eq.~(\ref{eq:Hdep}) should be 
sufficient to renormalize Eq.~(\ref{eq:LSeqCTCb}) including the
long-range Coulomb potential. 

\begin{figure}[ht]
\centerline{\includegraphics*[height=6cm]{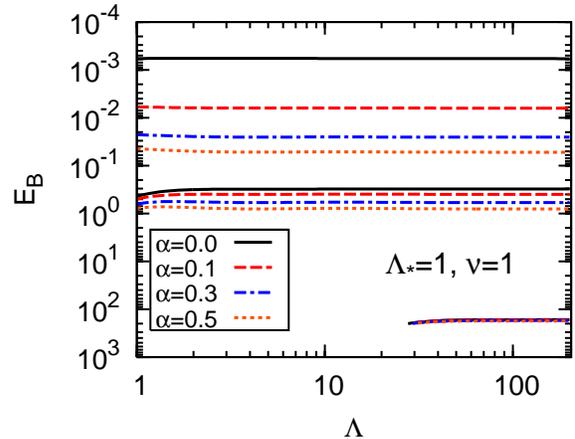}}
\caption{Binding energies for $\nu=\Lambda_*=1$ as function of the 
cutoff $\Lambda$ for $\alpha=0, 0.1, 0.3, 0.5$.}
\label{fig:plot1}
\end{figure}

In order to test this assumption, we calculate
the bound state spectrum using the counterterm 
for the pure $1/r^2$ potential from Eq.~(\ref{eq:Hdep}). 
The result is shown in Fig.~\ref{fig:plot1}. For simplicity, we 
take $\nu=\Lambda_*=1$ and show different values of $\alpha$.
For each energy level, the binding energies 
increase as the Coulomb strength is increased. It is evident that
the Coulomb potential influences the spectrum, but does
not destroy the cutoff independence of the binding energies. Clearly,
no additional counterterm is required. 

Note also that the 
shift in energy due to the Coulomb interaction depends on the exitation 
level --- the deepest bound states have the smallest shifts relative to the 
pure $1/r^2$ case, while the shallower ones have larger shifts and 
resemble more the Coulomb spectrum. This behavior can be understood by
comparing the relative strengths of the $1/r^2$ and $1/r$
potentials in coordinate space. There will always be a 
special distance $\bar r=|c|/\alpha=(1/4+\nu^2)/\alpha$ where
the two potentials have equal strength: 
$V_S(\bar r)=V_C(\bar r)$. 
The qualitative pattern of the energy levels can be understood by comparing
the binding energy of a given state, $E_B$, with the potential
energy at $\bar r$,
\beq
{\bar E}_{\rm pot}=|V_S(\bar r)+V_C(\bar r)|=\frac{2\alpha^2}{1/4+\nu^2}\,.
\label{eq:epot}
\eeq
The bound states with $E_B\gg {\bar E}_{\rm pot}$ are more
sensitive to shorter distances where $V_S$ dominates over $V_C$.
Therefore, the spectrum resembles the $1/r^2$ spectrum and is 
approximately scale invariant.
For $E_B\ll E_{\rm pot}$, on the other hand, the states are
more sensitive to larger distances and the spectrum is similar
to the Coulomb spectrum.
 
An alternative way of understanding this behavior is by looking at 
Eq.~(\ref{eq:LSeqCTCb}). We take one of the bound states $E_B=B_{n+1}$ 
with an eigensolution $\phi_{n+1}(p)$. The next deeper bound state, 
denoted by $B_n$, defines a variable $t_n$ ($>1$) via $t_n^2=B_n/B_{n+1}$. 
$B_n$ has an eigensolution $\phi_{n}(p)$ that satisfies
\begin{eqnarray}
\phi_n(p)&=&-\frac{1}{2\pi^2}\int_0^{\Lambda}\frac{dq\,q^2}{B_{n}\!+\!q^2}
\bigg\{2\pi^2c\bigg[\frac{\theta(p\!-\!q)}{p}+\frac{\theta(q\!-\!p)}{q}\bigg]
\nonumber\\ &&
+\frac{H_S(\Lambda/\Lambda_*)}{\Lambda}
-\frac{\pi\alpha}{pq}\ln\bigg[\frac{(p\!+\!q)^2}
{(p\!-\!q)^2}\bigg]\bigg\}\phi_{n}(q),
\label{eq:wavef1}
\end{eqnarray}
where we explicitly indicated the dependence of $H_S$ on $\Lambda$ and 
$\Lambda_*$. Rescaling the external momentum $p$ and the integration 
variable $q$ by $t_n$ yields 
\begin{eqnarray}
\phi_n(t_np)&=&
-\frac{1}{2\pi^2}\int_0^{\Lambda}\frac{dq\,q^2}{B_{n}/t_n^2\!+\!q^2}\times
\nonumber\\ &&
\bigg\{2\pi^2c\bigg[\frac{\theta(p\!-\!q)}{p}+\frac{\theta(q\!-\!p)}{q}\bigg]
+\frac{H_S(t_n\Lambda/\Lambda_*)}{\Lambda}
\nonumber\\ &&
-\frac{1}{t_n}\frac{\pi\alpha}{pq}\ln\bigg[\frac{(p\!+\!q)^2}
{(p\!-\!q)^2}\bigg]\bigg\}\phi_{n}(t_nq).
\label{eq:wavef2}
\end{eqnarray}
In addition, we set $\Lambda\to t_n\Lambda$ assuming cutoff 
independence, which is verified numerically. Writing 
$t_n=\exp(\pi/\nu+\delta_n)$ and using the log-periodicity of $H_S$, 
we conclude from
Eq.~(\ref{eq:wavef2}) that $B_{n+1}$ is also a bound state of 
a Hamiltonian that has a Coulomb potential weakened by $1/t_n$ 
and parameter $\Lambda_*$ multiplied by $e^{-\delta_n}$.
By induction, it follows that a deeper state $B_{n-k}$, with 
eigensolution $\phi_{n-k}(p)$, is given by 
$\sigma_k B_{n+1}=(\Pi_{j=0}^{k-1}t_{n-j})B_{n+1}$. Furthermore, $B_{n+1}$ 
is an eigenvalue, with eigensolution $\phi_{n-k}(\sigma_kp)$, of the 
Hamiltonian $H_0+V_S+V_C/\sigma_k$ with a parameter 
$\Lambda_*$ multiplied by $e^{-\Delta_k}=\exp(-\sum_{j=0}^{k-1}\delta_{n-j})$. 
Therefore, the solution $\phi_{n-k}(p)$ with eigenvalue $B_{n-k}$
is only weakly sensitive to the Coulomb 
potential. The same rationale applies to the solution $\phi_{n-k-1}(p)$ 
for the next deeper state $B_{n-k-1}$. One therefore expects that 
$\delta_{n-k-1}$ tends to zero and, consequently,  $t_{n-k-1}$ to the 
discrete scaling factor $e^{\pi/\nu}$. 

\begin{figure}[ht]
\centerline{\includegraphics*[height=6cm]{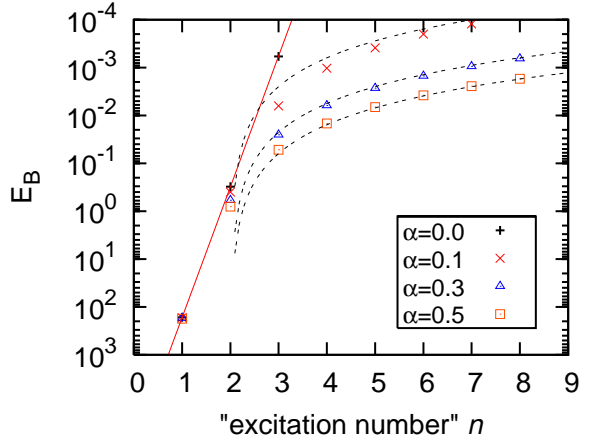}}
\caption{Binding energies for $\nu=\Lambda_*=1$ and
$\alpha=0, 0.1, 0.3, 0.5$ as function of the 
``exitation number'' $n$ (see text for details).}
\label{fig:plot1a}
\end{figure}
In order to support these conclusions, we plot in
Fig.~\ref{fig:plot1a} the binding energies as functions of 
the ``excitation number'' $n$, the latter being defined relative to the 
deepest bound state at $\Lambda=200$.\footnote{Note that this assignment
is not unique since new deep states appear as the cutoff is increased.
In the EFT framework, this is not a problem since all states
outside the range of validity of the EFT can be ignored.}
For $\nu=1$ and $\alpha=0.1, 0.3, 0.5$, one has 
${\bar E}_{\rm pot}=1.6\times 10^{-2}, 1.44\times 10^{-1}, 4.0\times 10^{-1}$, 
respectively.  
The exact scale invariance is broken for any finite value of $\alpha$.
The figure illustrates how the limit of exact discrete
scale invariance is approached as the states become deeper.
It confirms that for $E_{B}\gg {\bar E}_{\rm pot}$ the 
behavior is closer to the geometric $1/r^2$ spectrum
with $\ln E_B = {\rm const.}+n\times 2\pi/\nu$. This spectrum is
indicated by the solid straight line. 
For large $n$, where ${\bar E}_{B}\ll E_{\rm pot}$, 
the spectrum approaches the Coulomb spectrum. This is illustrated by 
the dotted lines which represent the Coulomb 
energies $\alpha^2/(4\tilde n^2)$, with the shifted excitation number 
$\tilde n=n-2$. 
This particular choice of $\tilde n$ is natural since the $n=3$ level 
is the closest to the $n=1$ level of the pure-Coulomb spectrum. 
In accordance with our
expectations, the Coulomb pattern is already observed at moderate $n$ 
for $\alpha=0.5$ and $0.3$, while it is achieved at larger $n$ for 
$\alpha=0.1$.

\begin{figure*}[ht]
\centerline{\includegraphics*[height=6cm]{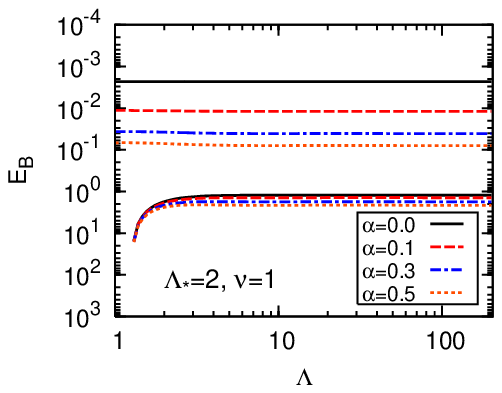}
\quad\includegraphics*[height=6cm]{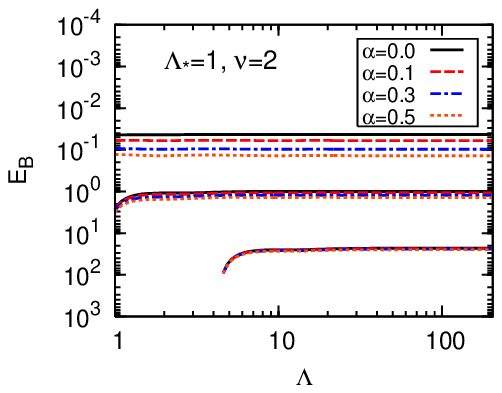}}
\caption{Bound state spectrum for $\Lambda_*=2$ and $\nu=1$
(left panel) and for $\Lambda_*=1$ and $\nu=2$
(right panel).}
\label{fig:plot3}
\end{figure*}

Next we study the dependence of the bound state spectrum on 
$\nu$ and $\Lambda_*$. In Fig.~\ref{fig:plot3}, we show the spectra for 
$\nu=1$ and $\Lambda_*=2$ (left panel) and for $\Lambda_*=1$ and $\nu=2$
(right panel). We first consider the left panel. Due to the
discrete scale invariance, a change of $\Lambda_*$ in the pure $1/r^2$ case 
modifies the values of the 
energies but preserves the geometric character of the spectrum,
i.e. the ratio of subsequent states is determined by the preferred
scaling factor squared $(e^{\pi/\nu})^2$. This behavior 
is manifest as a vertical displacement of the spectrum $\alpha=0$ relative 
to the one shown in Fig.~\ref{fig:plot1}.
A similar shift of the energies is also expected for $\alpha\neq 0$. 
However, the Coulomb interaction breaks the discrete scale invariance
and modifies the ratios between consecutive energy levels. 
The increase of $\Lambda_*$ provides more binding to 
the system, while the splitting among different values of $\alpha$ 
is reduced. This observation is in agreement with the finding in the
previous section: deeper bound states become less sensitive to the 
Coulomb part of the interaction, which is responsible for the 
different splittings. 
Next we consider the case $\Lambda_*=1$ and $\nu=2$ shown in the right
panel of Fig.~\ref{fig:plot3}. For a given value of $\alpha$,
the $1/r^2$ part of the interaction becomes relatively stronger
and the overall binding is increased. The influence of the 
long-range Coulomb part is decreased. For the pure $1/r^2$ case
$\alpha=0$, the discrete scaling factor is reduced to $e^{\pi/2}$ 
and the states move closer together. For the case $\alpha\neq 0$,
the spectrum has distinct features of the $1/r^2$ problem --- very little 
spread in energy among the different $\alpha$ considered, and a binding 
ratio very close to $(e^{\pi/2})^2$. 

\section{Perturbative Treatment of the Coulomb Part}
\label{sec:pert}

In the previous section we saw that the deepest bound states have 
small Coulomb corrections. In this section we calculate these 
energy shifts relative to the unperturbed 
$1/r^2$ solution in perturbation theory in order to test
our hypothesis.

It is convenient to start from the LS equation for the 
scattering state instead of the transition amplitude (\ref{eq:LSeqraw}). 
The former reads 
\begin{equation}
|\Psi_p^{(\pm)}\rangle=|\vec{p}\rangle+G_0^{(\pm)}(E)\left(\hat V_S+
\lambda\hat V_C\right)|\Psi_p^{(\pm)}\rangle\,,
\label{eq:LS4wf}
\end{equation}
where $|\Psi_p^{(+)}\rangle$ ($|\Psi_p^{(-)}\rangle$) is the initial 
(final) state, $G_0^{(\pm)}(E)$ is the free two-particle propagator, and 
$\lambda$ is a parameter expansion that will be set to 1 in the end. 
Expanding the scattering state and the binding energy in powers of 
a small parameter $\lambda$, 
\begin{eqnarray}
|\Psi_p^{(\pm)}\rangle&=&|\chi_p^{(\pm)}\rangle
+\lambda\,|\varphi_p^{(\pm)}\rangle+\cdots\,,
\nonumber\\[1mm]
E&=&E^{(0)}+\lambda\,E^{(1)}+\cdots\,,
\end{eqnarray}
one gets to leading order in $\lambda$ the LS equation for
the pure $1/r^2$ interaction $\hat V_S$:
\begin{equation}
|\chi_p^{(+)}\rangle=|\vec{p}\rangle+G_0^{(+)}(E^{(0)})\,\hat V_S\,
|\chi_p^{(+)}\rangle\,.
\label{eq:LSper0}
\end{equation}
Multiplication of this equation by $\langle\vec{p}'|\hat V_S$ 
gives the LS equation for $T_E(\vec{p},\vec{p}')\equiv 
\langle\vec{p}'|\hat V_S|\chi_p^{(+)}\rangle$
(cf.~Eq.~(\ref{eq:LSeqraw})). 
Collecting the terms linear in $\lambda$, we obtain
\begin{eqnarray}
|\varphi_p^{(+)}\rangle&=&\left[G_0^{(+)}(E^{(0)})\,\hat V_C
-E^{(1)}\,G_0^{(+)}(E^{(0)})^2\,\hat V_S\right]|\chi_p^{(+)}\rangle
\nonumber\\[1mm]&&
+G_0^{(+)}(E^{(0)})\,\hat V_S\,|\varphi_p^{(+)}\rangle\,.
\label{eq:LSper1}
\end{eqnarray}
This is an integral equation for the state $|\varphi_p^{(+)}\rangle$ 
with a driving term proportional to the unperturbed state 
$|\chi_p^{(+)}\rangle$. 
Since we are dealing with bound states, one can use the 
homogeneous version of Eq.~(\ref{eq:LSper0}) to eliminate the state 
$|\varphi_p^{(+)}\rangle$ from the above equation. This is achieved via 
multiplication from the left by $\langle\chi_{p'}|\hat V_S$ and yields 
\begin{eqnarray}
&&\langle\chi_{p'}^{(+)}|\hat V_S\,G_0^{(+)}(E^{(0)})\,
\hat V_C|\chi_p^{(+)}\rangle
\nonumber\\&&
-E^{(1)}\langle\chi_{p'}^{(+)}|\hat V_S\,G_0^{(+)}(E^{(0)})^2\,
\hat V_S|\chi_p^{(+)}\rangle=0\,.
\end{eqnarray}
We can again use Eq.~(\ref{eq:LSper0}) to express $E_B^{(1)}=-E^{(1)}$ in 
terms of the transition amplitude 
$T_E(\vec{p},\vec{p}')$ defined by Eq.~(\ref{eq:LSeqraw}). 
After integration over the angles, we obtain 
\begin{eqnarray}
E_B^{(1)}=-A/B
\end{eqnarray}
where
\begin{eqnarray}
\nonumber
A&=&\int_0^{\Lambda}dp\int_0^{\Lambda}dq\,
\frac{p^2\phi^{(0)}(p)}{E_B^{(0)}\!+\!p^2}\,v_C(p,q)\,
\frac{q^2\phi^{(0)}(q)}{E^{(0)}_B\!+\!q^2}\,,
\\[1mm]
B&=&2\pi^2\int_0^{\Lambda}dp\,
\frac{p^2\phi^{(0)}(p)^2}{(E_B^{(0)}\!+\!p^2)^2}\,,
\end{eqnarray}
and $\phi^{(0)}(p)$ is the solution of Eq.~(\ref{eq:LSeqCT}).

\begin{figure*}[htb]
\centerline{\includegraphics*[height=6cm]{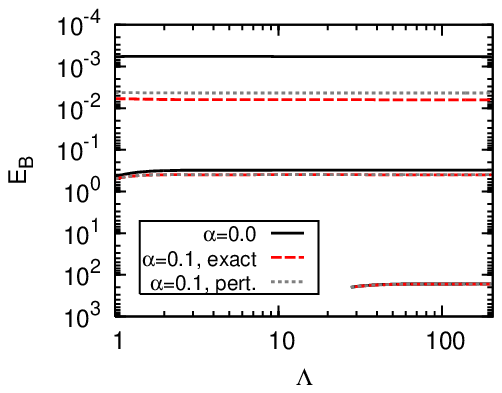}
\quad\includegraphics*[height=6cm]{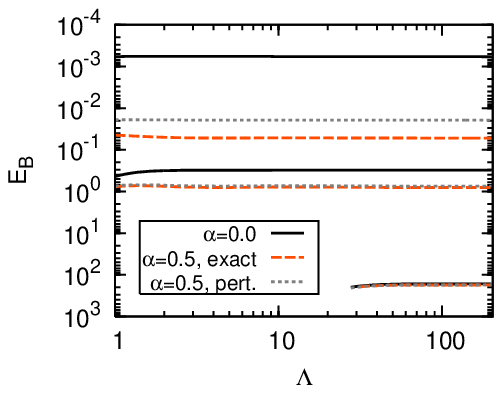}}
\caption{Bound state spectrum with a perturbative Coulomb treatment for 
$\alpha=0.1$ (left panel) and $\alpha=0.5$ (right panel), compared against 
the exact and $\alpha=0$ cases. The parameters $\Lambda_{*}$ and $\nu$ 
were set to 1.}
\label{fig:plot4}
\end{figure*}

A comparison of the exact binding energies with the perturbative
results is given in Fig.~\ref{fig:plot4}.
We show the perturbative Coulomb binding energies 
for $\alpha=0.1$ (left panel) and $0.5$ (right panel),
compared with exact energies and the energies for the pure $1/r^2$ 
case. On one hand one clearly sees that the perturbative treatment works 
quite well for the two deepest bound states, where the effect of the 
Coulomb interaction is expected to be small. This is true even for 
a relatively strong Coulomb potential with $\alpha=0.5$.
The shallowest state in Fig.~\ref{fig:plot4}, on the other hand,
cannot be described by perturbation theory in the Coulomb potential.
In this case, we no longer have $E_B\gg {\bar E}_{\rm pot}$
(cf.~Eq.~(\ref{eq:epot})) and the Coulomb effects are large.
Indeed, the perturbative treatment of the Coulomb interaction gives 
$4.4\times 10^{-3}$ ($1.9\times 10^{-2}$) compared to the exact value 
$6.3\times 10^{-3}$ ($5.2\times 10^{-2}$), for $\alpha=0.1$ ($\alpha=0.5$).
These results clearly support our hypothesis from the previous section.

We should point out that a perturbative treatment of the $1/r^2$ 
potential relative to the Coulomb potential to calculate the shallower 
states is not possible. This is due to the singular nature of
the $1/r^2$ potential for real values of $\nu$ in Eq.~(\ref{eq:defpot})
which we consider here. We have verified explicitly that
if the  $1/r^2$ potential is not singular (corresponding to 
imaginary $\nu$) the perturbative treatment works quite well.
The latter case, however, 
corresponds to a situation where the Schr\"odinger equation has 
a unique solution and limit cycles are absent.

\section{Summary and Conclusions}
\label{sec:conc}

In this work, we have investigated the modification of limit cycles
and discrete scale invariance by the presence of a long-range
interaction. As a specific example, we have considered the 
quantum mechanical inverse square potential supplemented by
an attractive long-range Coulomb interaction. We have focused
on the bound state properties of this model system.

Our study of the cutoff dependence of the binding energies 
shows that no additional counterterm is required
for renormalization when the Coulomb potential is added. The 
counterterm that renormalizes the inverse square potential alone
is sufficient to renormalize the full problem.

In the presence of the Coulomb potential, the counterterm can no 
longer be obtained analytically.
We have calculated the counterterm numerically by fixing  one
of the bound state energies. All other binding energies are then
independent of the ultraviolet cutoff $\Lambda$.
This procedure has been carried out for several values of Coulomb
strength parameter $\alpha$. The counterterm was confirmed 
to be a log-periodic function with discontinuities.
Its $\Lambda$-dependence is the same as for the pure inverse
square potential but shifted 
along the $\Lambda$-axis. Such a translation corresponds to a finite 
renormalization of the counterterm parameter $\Lambda_*$.

The discrete scale invariance of the inverse square potential is 
broken by the Coulomb potential. 
We have investigated the deviations from discrete scaling symmetry for 
different strengths of the Coulomb potential. For highly excited 
states, the long-distance Coulomb tail dominates the 
dynamics and the levels tend towards a Coulomb spectrum. 
The deepest bound states, however, are mostly sensitive to the 
short-range $1/r^2$ potential and show an approximate scaling symmetry.
The spectra obtained for various Coulomb strengths 
and values for $\Lambda_{*}$ and $\nu$ were studied in 
detail. Due to the breaking of scaling symmetry, 
the ratio between consecutive bound state energies (the discrete scaling 
factor) is no longer a constant. The splittings of the energy
levels depend on the magnitude of the binding and the Coulomb 
strength.

To verify our conclusions, we have studied the behavior of the deep 
bound states in perturbation theory. 
We have derived an expression for the energy shift relative to the 
$1/r^2$ spectrum, treating the Coulomb interaction in first-order 
perturbation theory. We have shown that the perturbative
expression is valid for bound states that satisfy 
$E_B\gg {\bar E}_{\rm pot}$, where ${\bar E}_{\rm pot}$ is the potential 
energy at the distance $\bar r$ where both interactions have
equal strength.

After the general features of the breaking of discrete scale
invariance are understood for this example, we are in 
the position to study more realistic systems.  Our results 
could be useful for the study of nuclear cluster states in the halo 
EFT \cite{BHvK1,BHvK2,BRvK,Higa:2008dn}.
In Ref.~\cite{Higa:2008dn}, e.g., a power counting scenario for
the $\alpha\alpha$ system was formulated. According to this scenario
the ${}^8$Be system would exhibit conformal invariance at leading order
and ${}^{12}$C would display an exact Efimov spectrum. 
These exact features are broken by the Coulomb interaction but some 
remnants of this behavior are manifest in the experimental spectra,
such as the shallowness of the ${}^8$Be $0^+$ resonance.
The $^{12}$C Hoyle state would then be
a remnant of an Efimov state that appears in the limit of large
scattering length.
An application of the power counting scenario \cite{Higa:2008dn}
to the triple-$\alpha$ system remains to be carried out.
Our calculation provides a first step towards the understanding
of the breaking of discrete scale invariance in these systems. 
Additional expansions such as a strong
coupling expansion for the Coulomb interaction \cite{Higa:2008dn}
might be useful and deserve further study.

\section*{Acknowledgments}
This research was supported in part by the
BMBF under contract number 06BN411.

\appendix
\section{Treatment of the Coulomb Divergence}\label{app}
\label{sec:CBdiv}

In this Appendix, we describe our method to treat the Coulomb
singularity in Eq.~(\ref{eq:LSeqCTCb}).

The integral equation for bound states (\ref{eq:LSeqCTCb}) is conveniently 
rewritten as\footnote{We do not explicitly display 
the counterterm $H_S(\Lambda)$
since it is irrelevant for the Coulomb divergence problem.} 
\begin{eqnarray}
\Big[1\!-\!F(p)\Big]\,\phi_E(p)&=&\frac{1}{2\pi^2}\int_0^{\Lambda}
\frac{dq\,q^2}{E\!-\!q^2}\Big[v_S(p,q)\nonumber\\
&&+v_C(p,q)\Big]\,
\phi_E (q)-F(p)\,\phi_E(p)\,,
\nonumber\\
&& \label{eq:inteq02}
\end{eqnarray}
where the function $F(p)$ was introduced on both sides of the 
equation to cancel the Coulomb divergence in the diagonal terms. 
Following Ref.~\cite{Coultrick}, one finds that a suitable choice for 
$F(p)$ is given by 
\begin{eqnarray}
F(p)&=&\int_{\Lambda}\frac{d^3q}{(2\pi)^3}\,\frac{(E-p^2)}{(E-q^2)^2}\,
V_C(|\vec{p}-\vec{p}'|)\nonumber\\
&=&
\frac{1}{2\pi^2}\int_0^{\Lambda}dq\,q^2\,\frac{(E-p^2)}{(E-q^2)^2}\;v_C(p,q)\,.
\label{eq:trick}
\end{eqnarray}
The integral can be evaluated analytically: 
\begin{eqnarray}
F(p)&=&\frac{\alpha\,(E-p^2)}{(E\!-\!p^2\!+\!\gamma^2)^2+4p^2\gamma^2}
\nonumber\\
&& \times \Bigg\{
\frac{(E\!-\!p^2\!-\!\gamma^2)}{\pi\sqrt{-E}}\,
\arctan\left(\frac{\Lambda}{\sqrt{-E}}\right)
\nonumber\\[1mm]&&
+\frac{\gamma}{\pi}\left[\arctan\left(\frac{\Lambda+p}{\gamma}\right)
+\arctan\left(\frac{\Lambda-p}{\gamma}\right)\right]
\nonumber\\[1mm]&&
-\frac{1}{4\pi p(E\!-\!\Lambda^2)}\bigg[(E\!-\!p^2\!+\!\gamma^2)
(\Lambda^2\!-\!p^2\!+\!\gamma^2)\nonumber\\
&&+4p^2\gamma^2\bigg]
\ln\left[\frac{(p\!+\!\Lambda)^2+\gamma^2}{(p\!-\!\Lambda)^2+\gamma^2}\right]
\Bigg\}
\label{eq:trick01}
\\[1mm]&\stackrel{\Lambda\to\infty}{=}&
\frac{\alpha\,(E-p^2)}{(E-p^2+\gamma^2)^2+4p^2\gamma^2}
\left[\frac{E-p^2-\gamma^2}{2\sqrt{-E}}+\gamma\right]\,.
\nonumber\\ &&
\label{eq:trick02}
\end{eqnarray}

Using Eq.~(\ref{eq:trick}) on the r.h.s. and either Eqs.~(\ref{eq:trick01}) 
or (\ref{eq:trick02}) on the l.h.s. of Eq.~(\ref{eq:inteq02}) is enough to 
eliminate the divergence problem. As a numerical check, we set $v_S\to 0$ 
and obtained very stable and accurate values for the Coulomb spectrum. 
For the values of the Coulomb strength $\alpha$ considered in this work, 
we also observed cutoff independence except at lower values 
($\Lambda\lesssim 1$), where results using (\ref{eq:trick01}) or 
(\ref{eq:trick02}) start to deviate by a few percent.


\end{document}